\documentclass[aps,prl,twocolumn,showpacs]{revtex4-1}
\usepackage{epsfig}
\usepackage{graphicx}
\usepackage{amsmath}
\usepackage{bm}

\begin{document}

\DeclareGraphicsExtensions{.eps,.EPS}

\title{Spin mixing and protection of ferromagnetism in a spinor dipolar condensate}
\author{S. Lepoutre$^{1,2}$, K. Kechadi$^{1,2}$, B. Naylor$^{1,2}$, B. Zhu$^{3}$, L. Gabardos$^{1,2}$, L. Isaev$^{3}$, P. Pedri$^{1,2}$, A. M. Rey$^{3}$, L. Vernac$^{1,2}$, B. Laburthe-Tolra$^{2,1}$.}

\affiliation{$^{1}$\,Universit\'e Paris 13, Sorbonne Paris Cit\'e, Laboratoire de Physique des Lasers, F-93430, Villetaneuse, France\\
$^{2}$\,CNRS, UMR 7538, LPL, F-93430, Villetaneuse, France\\
$^{3}$\, JILA, NIST and Department of Physics, University of Colorado, Boulder, USA}

\begin{abstract}
We study spin mixing dynamics in a chromium dipolar Bose-Einstein Condensate, after tilting the atomic spins by an angle $\theta$ with respect to the magnetic field. Spin mixing is triggered by dipolar coupling, but, once dynamics has started, it is mostly driven by contact interactions. For the particular case $\theta=\pi/2$, an external spin-orbit coupling term induced by a magnetic gradient is required to enable the dynamics. Then the initial ferromagnetic character of the gas is locally preserved, an unexpected feature that we attribute to large spin-dependent contact interactions.

\end{abstract}
\date{\today}
\maketitle

For spin systems, insight on their collective behaviour can be gained from observations following a rotation of the individual spins initially oriented along the external magnetic field. Deviation from an overall precession of the spins at the Larmor frequency may reveal interparticle interactions. This generic problem is encountered for example in Nuclear Magnetic Resonance, where dipole-dipole interactions (DDIs) is a source of decoherence \cite{BookNMR}, and thus have to be reduced to obtain long coherence lifetimes of nuclear spins \cite{Waugh67}. Similar cancellation of DDIs has also been demonstrated in the case of electrons in semiconductors \cite{Si2012}, or for impurity centers in a solid \cite{Hanson2008}.

\begin{figure}
\centering
\includegraphics[width= 3 in]{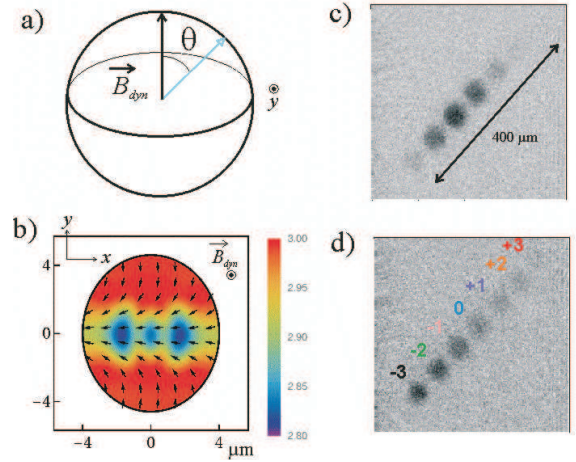}
\caption{\setlength{\baselineskip}{6pt} {\protect\scriptsize Principle of the experiment. a) The spins of atoms in a polarized $s=3$ $^{52}$Cr BEC are rotated at $t=0$, and make an angle $\theta$ with respect to the external magnetic field, $\textbf{\textit{B}}_{\rm dyn}$. b) After a variable time $t_{\rm dyn}$ spins point in different directions (shown by arrows) but have a length almost constant (see color code). The figure is a result of our Gross Pitaevskii simulation for $t_{\rm dyn}=5$ ms. Right: Absorption imaging after a Stern Gerlach separation allows to measure populations in the seven spin components. Pictures show the seven separated clouds for $\theta=\pi/2$ and the magnetic field configuration of Fig. \ref{DynPiover2}. c) $t_{\rm dyn}=0.1$ ms. d) $t_{\rm dyn}=5$ ms.}}
\label{Principle}
\end{figure}

On the contrary, interaction-induced modification of the mere precession can be desired, as it leads to interesting phenomena. For example DDIs between ultracold molecules in an optical lattice were evidenced this way \cite{Ye2013}. For spinor quantum gases \cite{DSKUeda}, i.e. quantum degenerate gases with a spin degrees of freedom, tilting the spins by an angle $\theta$ with respect to the magnetic field (see Fig. \ref{Principle}a)) has lead to observation of spontaneous pattern formation due to instabilities driven by DDIs \cite{DSK2008,Hirano2014} or antiferromagnetic contact interaction \cite{Sengstock2010}. Dipolar spin systems could display beyond mean field physics for angles $\theta$ close to $\pi/2$ \cite{Hazzard2013}. Here we rotate the spins of a $s=3$ $^{52}$Cr BEC, and study how spin dynamics develops as an interplay between  contact interactions, DDIs, and magnetic field gradients (MGs). One unexpected outcome is that strong spin-dependent contact interactions favor the persistence of ferromagnetic textures (see Fig. \ref{Principle}b)), and seem to slow down beyond mean-field effects.

Compared to alkaline spinor gases, a $^{52}$Cr BEC offers two key differences: spin dependent contact interactions are significantly larger \cite{Naylor2016}, and DDIs are 36 times larger. After rotation from an initially polarized BEC in $s=3, m_s=-3$, atoms are in a stretched (ferromagnetic) state corresponding to a well-defined molecular potential. Starting from this initial state, no spin dynamics can develop under the influence of contact interactions (in contrast with \cite{dePaz2016}, where a non-trivial initial state was prepared), which display $SU(2)$ symmetry and preserve the total spin. On the other hand, DDIs or MGs, which convey spin-orbit coupling, can trigger spin mixing. For example, in the mean-field approximation, the dipoles which precess around the external magnetic field create an effective dipolar magnetic field \cite{Kawaguchi2007} proportional to the collective magnetization along the  quantization axis: DDIs result in precession of the spins around this dipolar field and trigger spin dynamics, unless $\theta = \pi /2$ in which case the initial collective magnetization vanishes.

Experimentally, we find that DDIs do trigger spin dynamics, even when we best suppress MGs, for $\theta \neq \pi /2$. However, spin dynamics is strongly suppressed for $\theta = \pi /2$ without MGs. For $\theta \approx \pi /2$, we recover spin dynamics by applying MGs. In that case our numerical simulations show that dynamics preserves the local spin length. We attribute this unexpected effect to large spin-dependent contact interactions: depolarization is inhibited by an energy gap $\propto 4 \pi \hbar^2 n (a_6-a_4)/M$ (with $M$ the mass, $n$ the density, and $a_S$ the scattering length associated to the molecular potential $S$, see full Hamiltonian below), favouring persistence of locally fully magnetized classical states.

The starting point of our experiments is a polarized $^{52}$Cr BEC produced in a crossed dipole trap \cite{Naylor2016}, with typically $4\times10^4$ atoms polarized in the minimal Zeeman energy state $m_s=-3$. Trap frequencies are $\omega_{x,y,z}=2\pi\times(298,245,210)$ Hz, with $5\%$ uncertainty, $Oy$ being the vertical axis. The atomic spins are aligned along an external magnetic field noted $\textbf{\textit{B}}_{\rm dyn}$, whose average amplitude $B_{0}$ is in the $150-200$ mG range, and whose average direction $\hat{u}_B$ is in the horizontal $xz$ plane. $\textbf{\textit{B}}_{\rm dyn}$ is maintained constant during the whole dynamics, with a reproducibility of 1 mG. The dynamics is initiated at $t=0$ by a resonant radio-frequency pulse, at the Larmor frequency $f_L=g\mu_B B_0$ ($g=2$ is the Land\'e factor, $\mu_B$ the Bohr magneton), and a Rabi frequency equal to $50$ kHz; it rotates all spins by an angle $\theta$, so that $\textbf{\textit{s}}_{t=0^+}=\cos(\theta)\hat{u}_B+\sin(\theta)\hat{u}_{\perp}$, ($\hat{u}_{\perp}\perp (\hat{u}_{B},\hat{u}_{y})$, see Fig. \ref{Principle}a)). We let the system evolve for a duration $t_{\rm dyn}$. We then switch the optical trap off and spatially separate the $7$ spin components during a time of flight of $5$ ms, using a Stern Gerlach (SG) technique (see Fig. \ref{Principle}c)d)). We adiabatically change the magnetic field so that spin populations measured after SG correspond to a projective measurement along $\hat{u}_B$ \cite{SuppMat}. From absorption pictures as those shown in Fig. \ref{Principle}c)d), we obtain populations $N_{m_s}$ of all seven spin components (for details on the detectivity calibration of absorption images see \cite{SuppMat}). We then compute the total population $N_{tot}=\sum_{m_s}{N_{m_s}}$ and fractional populations $p_{m_s}=N_{m_s}/N_{tot}$. We also measured carefully the 3D spatial dependance of $\textbf{\textit{B}}_{\rm dyn}$, and obtained $\left|\textbf{\textit{B}}_{\rm dyn}\right|=B_0+\textbf{\textit{b}}. \textbf{\textit{r}}$, with $\textbf{\textit{b}}$ determining the  magnetic field gradient \cite{SuppMat}.

\begin{figure}
\centering
\includegraphics[width= 3.5 in]{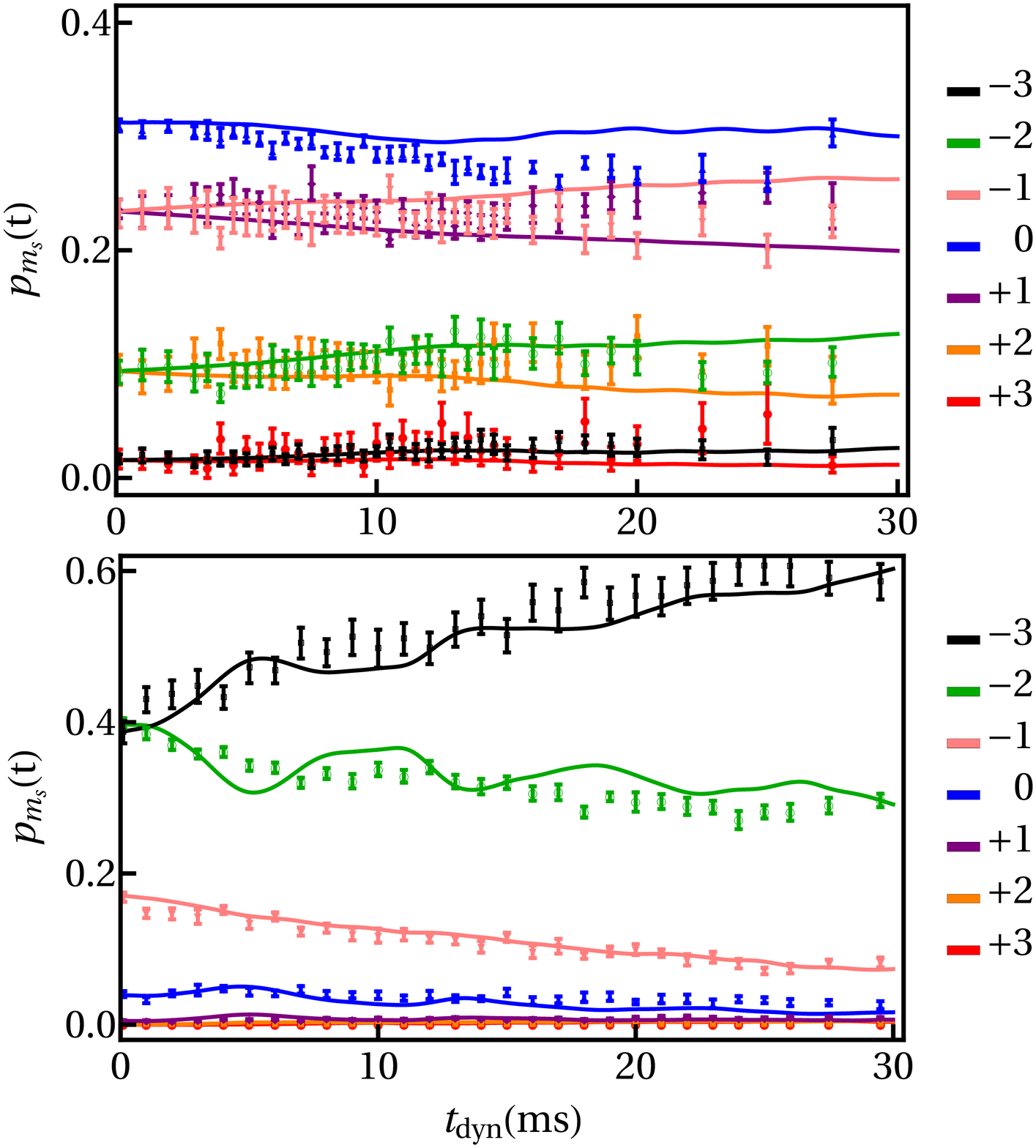}
\caption{\setlength{\baselineskip}{6pt} {\protect\scriptsize Temporal evolutions for $B_0=189$ mG, $\hat{u}_{B}=\cos\left(34\times \pi /180\right)\hat{u}_{x}+\sin\left(34\times \pi /180\right)\hat{u}_{z}$, $\left(b_x,b_y,b_z\right)=\left(3.5\pm15,3.6\pm7,1.8\pm17\right)$ mG.cm$^{-1}$. a) $\theta = \pi /2$. b) $\theta = \pi /4$. We show experimental data (points) and result of our Gross Pitaevskii simulations without free parameters (solid line) for the seven fractional populations. Error bars take into account both statistical and systematics uncertainties.}}
\label{DynPiover2and4}
\end{figure}

Figure \ref{DynPiover2and4} displays temporal evolution of $p_{m_s}$, for a $\textbf{\textit{B}}_{\rm dyn}$ configuration corresponding to our best cancellation of MGs. There is almost no spin mixing dynamics in the case $\theta=\pi /2$ as shown by Fig. \ref{DynPiover2and4} a). In absence of MGs, this is expected from the mean-field point of view \cite{Kawaguchi2007}. On the other hand, spin models do predict beyond mean field dynamics at $\theta=\pi /2$, over a typical timescale given by the quadratic average of DDIs \cite{Hazzard2013}, $\simeq 10$ ms for our atomic distribution. Beyond mean field simulations of our BEC system based on the Truncated Wigner Approximation \cite{Blakie2008} indicate that the gap associated to spin-dependent short range interactions delay the onset of beyond mean field corrections, and can account for the small variation in populations and corresponding deviation of the data  from the mean field dynamic observed at
long times. A more systematic study remains to be done in order to verify whether other systematic effects (e.g. associated to the existence of a small tensor light shift, see below) are at play.

In contrast, spin mixing dynamics is obtained for a rotation by $\theta=\pi /4$ when MGs are best cancelled (see Fig. \ref{DynPiover2and4} b)). The occurrence of spin dynamics in absence of MGs is an experimental signature of DDIs, which are the only ones to break spin rotational symmetry in our setup. In contrast no spin dynamics would arise for $SU(2)$ spin models like the Heisenberg spin model. The demonstration of this genuine dipolar dynamics in a BEC is the first main result of this paper.

\begin{figure}
\centering
\includegraphics[width= 3.5 in]{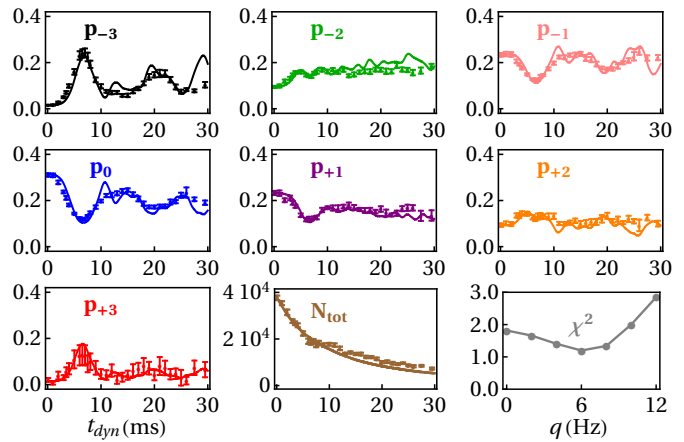}
\caption{\setlength{\baselineskip}{6pt} {\protect\scriptsize Temporal evolution following an initial rotation by an angle $\theta = \pi /2$, for $B_0=170$ mG,
$\hat{u}_{B}=\hat{u}_{z}$, $\left(b_x,b_y,b_z\right)=\left(1.6\pm7,44.6\pm7,5.7\pm8\right)$ mG.cm$^{-1}$. We show experimental data (points) and results of Gross Pitaevskii simulations (full lines), for the seven fractional populations, and for the total number of atoms (bottom center). Error bars take into account both statistical and systematics uncertainties. Bottom right: influence of a non zero quadratic effect. The $\chi ^2$ criteria, evaluated over the full spin populations data, is plotted as a function of $q$.
}}
\label{DynPiover2}
\end{figure}

Figure \ref{DynPiover2} shows evolution of $p_{m_s}$ for $\theta=\pi /2$ in a case where MGs are not minimized, contrarily to Fig. \ref{DynPiover2and4} a). Then a significant dynamics is observed. We find that the general trend is that larger MGs lead to faster spin dynamics (see \cite{SuppMat}). Besides we find a significant decrease of $N_{tot}$ in a few ms, which we attribute to dipolar relaxation collisions \cite{PasquiouPRA}. Losses due to dipolar relaxation are more important at $\theta=\pi /2$ compared to $\theta=\pi /4$ due to the larger spin rotation.

The ensemble of results shown in figures \ref{DynPiover2and4} and \ref{DynPiover2} demonstrate the sensitivity of the dynamics not only to the external magnetic field configuration, but as well to the initial preparation (angle $\theta$). In order to understand the influence of these different parameters, and the role played by the different interactions, we have developed a three dimensional spinorial Gross-Pitaevskii (GP) simulation. At the mean field level the Hamiltonian of the system is given by:

\begin{eqnarray}
  && H=\int d^3 {\bf r}\Big( {\bf \Psi}^\dagger \hat{H}_0{\bf \Psi}+ \mu_B g \left|\textbf{\textit{B}}_{\rm dyn}\right| S^Z({\bf{r}}) +\frac{c_0}{2}|n({\bf{r}})|^2\Big) \nonumber
   \\&&+\int d^3{\bf r}\Big(  \frac{c_1}{2}| {\bf{S}}({\bf{r}})|^2+\frac{c_2}{2}|A_{00}({\bf{r}})|^2+ \frac{c_3}{2} \sum_{M=-2}^{2} \vert A_{2M} \vert^2 \Big)\nonumber
   \\&&+\frac{c_{dd}}{2}\int  d^3{\bf r}  d^3{\bf r'}\frac{1-3 (\hat{e}\cdot \hat{\textit{u}}_B)^2}{|{\bf r}- {\bf r'}|^3}
\Big[{{S}^Z}({\bf{r}}) {{S}^Z}({\bf{r'}})\nonumber
\\&&-\frac{1}{2} \Big({{S}^X}({\bf{r}}){{S}^X}({\bf{r'}}) + {{S}^Y}({\bf{r}}){{S}^Y}({\bf{r'}})\Big)\Big] -i\Gamma
\label{Hamiltonian}
\end{eqnarray}
where  ${\bf \Psi}$ is a seven-component spinor describing the condensate field. The spin density vector is ${\bf {S}}({\bf{r}})={\bf \Psi}^\dagger ({\bf{r}})\cdot{\bf s}\cdot {\bf \Psi}({\bf{r}})$ with ${\bf s}=\{s^X,s^Y,s^Z\}$ spin-3 matrices. $\hat{H}_0=-\frac{\hbar^2}{2M}\nabla^2+ V_{\rm trap}({\bf r})$ is the single particle Hamiltonian, with $V_{\rm trap}({\bf r})$  the spin-independent harmonic trapping potential.

The term proportional to $c_0$ describes the contact density interaction, with $n({\bf{r}})={\bf \Psi}^\dagger({\bf{r}}) \cdot {\bf \Psi}({\bf{r}})$  the condensate density. The terms proportional to $c_1,c_2,c_3$ describe the contact  spin
dependent interactions  \cite{Uedareport}, with  $A_{SM}(\textbf{\textit{r}}) = \sum_{mm' = -3}^{3} \left\langle S,M \vert 3,m;3,m' \right\rangle \Psi_{m}(\textbf{\textit{r}}) \Psi_{m'}(\textbf{\textit{r}})
$.  For a $s=3$ spinor, $c_0 = g_c(9 a_4 + 2 a_6)/11$, $c_1 = g_c(a_6 - a_4)/11$, $c_2 = g_c(11 a_0 - 21 a_4 + 10 a_6)/11$, $c_3 = g_c(11 a_2 - 18 a_4 + 7 a_6)/11$, with $g_c=4\pi\hbar^2/M$. For $^{52}$Cr, we use $c_0 = 71$ $g_c a_B$, $c_1 = 3$ $g_c a_B$; $c_2 = -15$ $g_c a_B$, $c_3 = -46$ $g_c a_B$, with $a_B$ the Bohr radius \cite{Werner2005,PasquiouPRA,dePaz2014}.

The term proportional to   $c_{dd}= \mu_0 (g\mu_B)^2/(4\pi)$ (with  $\mu_0$ the magnetic permeability of vacuum, and $\hat{e} = ({\bf{r}} - {\bf{r'}})/|{\bf{r}} -{\bf{r'}}|$) is the secular Hamiltonian of DDIs which conserves magnetization along the magnetic field. In addition, magnetization non-conserving terms of DDIs give rise to the observed atom losses, which we take into account by adding the imaginary term $i\Gamma$, following \cite{BECbook}.

The Hamiltonian of eq.(\ref{Hamiltonian}) can be enriched by a quadratic term $q (S^Z)^2$; for Cr, it corresponds to the (small) tensorial light shift induced by the 1075 nm laser creating the optical trap. We show in Fig. \ref{DynPiover2} that $q\simeq6$ Hz leads to slightly better agreement with the data.

\begin{figure}
\centering
\includegraphics[width= 3.0 in]{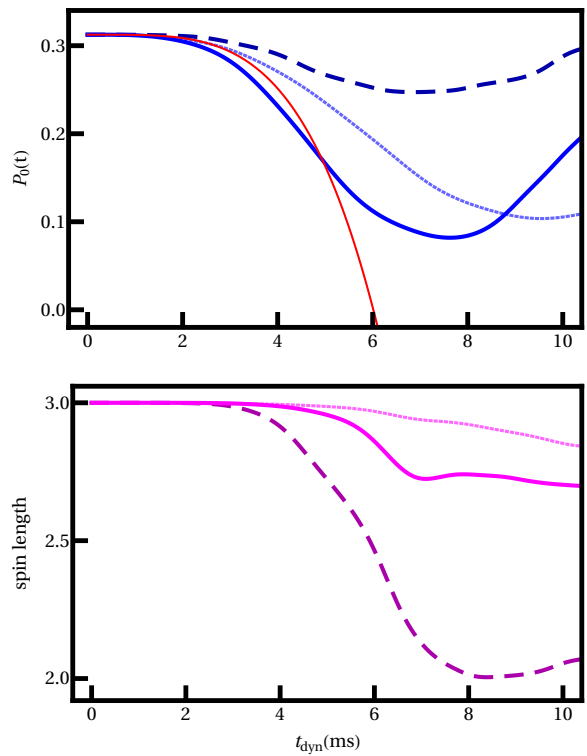}
\caption{\setlength{\baselineskip}{6pt} {\protect\scriptsize Results of numerical simulations for the same magnetic field configuration as in Fig. \ref{DynPiover2}. We compare results for the real case (thick full lines) to the case where contact spin dependent (dashed lines) or dipolar interactions (dotted lines) are suppressed (see text). a): fractional populations in $m_s=0$. The (red) thin line corresponds to our model at short time (see eq.(\ref{ModelMG})). b): spin length integrated over all cloud.}}
\label{fig4}
\end{figure}

The GP simulations compare well to experimental data (see Figs. \ref{DynPiover2and4}, \ref{DynPiover2}), with no adjustable parameter ($q=0$). Simulations thus allow us to probe the relative influence of the different interactions. We show on Fig. \ref{fig4}a) what is the expected dynamics if DDIs or contact spin dependent interactions are neglected; in the latter case spin independent interactions are kept constant by artificially setting $a_0=a_2=a_4=a_6=(9a_4+2a_6)_{\text{real}}/11)$. The striking differences seen for the different cases   show the prominent role played by contact spin exchange processes. In this example  $\theta=\pi/2$, but  even in the case of
$\theta=\pi/4$, where DDIs are instrumental in triggering spin
dynamics (see Fig. \ref{DynPiover2and4}b)), similar conclusions hold.

Furthermore, simulations confirm that in presence of small MGs spin mixing develops for $\theta=\pi/4$ (see Fig. \ref{DynPiover2and4} b)); and that, on the contrary, spin mixing dynamics is much reduced for $\theta=\pi/2$ (see Fig. \ref{DynPiover2and4} a)), unless a MG is applied (see Fig. \ref{DynPiover2}). This was discussed in \cite{Hirano2014} for the particular case $\textbf{\textit{b}}//\hat{u}_B$; our simulations indicate that MGs with $\textbf{\textit{b}}\perp \hat{u}_B$ are even more efficient to trigger dynamics. Therefore MGs play a key role to trigger spin mixing dynamics in our system (as opposed to e.g. \cite{Santos2010}, where it is shown that MGs can suppress spin exchange processes).

Just after rotation, all spins are aligned and the sample is ferromagnetic. One striking observation in our simulations is that a local ferromagnetic character is maintained while dynamics proceeds. Indeed, the local spin length $\Pi(\textbf{\textit{r}})=\left|\bf{S}(\textbf{\textit{r}})\right|/n(\textbf{\textit{r}})$ ($0\leq \Pi(\textbf{\textit{r}})\leq 3$) integrated over the cloud remain (almost) maximal, even after a large dynamics has developed (see Fig. \ref{fig4}b)). This protection of local ferromagnetism, which would be even more pronounced in absence of DDIs (see Fig.\ref{fig4} b)), comes as a surprise since $a_6>a_4$ energetically favors depolarization in the $^{52}$Cr BEC \cite{Pasquiou2011}.

To understand this effect, we have solved the GP equation for a homogeneous BEC in presence of MGs. We find that the initial ferromagnetic character of the BEC is protected by spin exchange contact interactions, which provide self-rephasing of the spinor components, similar to the one observed in atomic clocks \cite{Rosenbusch2010}. This mechanism ensures protection of polarization as long as it overcomes the phase scrambling between the different $m_s$ components due to increasing kinetic energy. For a homogeneous gas of density $n$ after an evolution time $T$, this requires $\left(m_s g \mu_B b \right)^2 T^2 /(2M)<<c_1n$ \cite{SuppMat} (and provided incoherent scattering remains negligible \cite{Piechon2009}). For a trap this criterion must be fulfilled at $T=\pi/(2\omega)$ when kinetic energy is maximum.  In our experiment the ratio between these two energies is about $1/40$.

Taking the phenomenological assumption that the local spinor remains ferromagnetic, we derive the following evolution of the fractional populations \cite{SuppMat}, assuming that the system is described by the ferromagnetic hydrodynamics equations \cite{Kudo2010}  (which underlines a connection to the physics of ferrofluids), and taking an initial Gaussian ansatz of $1/e^2$ radius $R$:

\begin{equation}
\frac{p_{m_s}(t)}{p_{m_s} (0)}= 1 + \frac{1}{2}\left(\frac{g \mu_B b}{M R}\right)^2 \left( m_s^2- \sum_{m_{s'}}  m_{s'}^2 p_{m_s'}(0) \right) t^4
\label{ModelMG}
\end{equation}
In this picture, spin exchange processes act to counterbalance the modification of the local fractional populations due to separation between spin components induced by MGs. Then eq.(\ref{ModelMG}) provides a time scale $(MR/g \mu_B b)^{1/2}$ for the spin dynamics. We find excellent agreement with simulations at short time when only contact interactions are taken into account (see Fig. \ref{fig4}a), and \cite{SuppMat}). We stress that this model is independent of interactions, which are in practice adiabatically eliminated. Large enough spin-dependent interactions thus convey metastability to ferromagnetism, whether the spinor ground state is ferromagnetic, or not (e.g. $^{52}$Cr is expected to have a cyclic ground state \cite{dePaz2014}). This finding is the second main result of our paper.

While the spins remain (almost) locally polarized, our GP simulations show that spin textures grow, which mostly corresponds to $\bf {S}(\textbf{\textit{r}})$ pointing in different directions (see Fig. \ref{Principle}b)). In contrast to \cite{Hirano2014} in situ spin structures are strongly modified after time of flight in our case, and their study goes beyond the scope of this paper. These textures result from an interplay between contact and dipolar interactions. Spin textures of dipolar origin can be created at a rate of order \cite{Kawaguchi2007} $\Gamma_{dd}=(3n \text{c}_{\text{{\tiny dd}}})/\hbar=2505$ s$^{-1}$, with $n=2.5\times10^{20}$m$^{-3}$. On the other hand, $\Gamma_{S,S'}=4\pi \hbar n (a_S-a_{S'})/m$, $\Gamma_{4,6}=8\times10^3$ s$^{-1}$. These estimates confirm the interplay between contact and dipolar interactions. In contrast to $F=1$ $^{87}$Rb \cite{DSK2008}, $^{52}$Cr does not behave as a purely dipolar quantum gas.

In conclusion, we have investigated the spin mixing dynamics for a $s=3$ spinor BEC following a rotation of the individual spins, initially aligned along the magnetic field. We have observed that while DDIs can trigger dynamics for $\theta \neq\pi/2$, MGs are necessary for $\theta=\pi/2$. For this case, we have demonstrated the occurrence of an original scenario, in which strong spin dependent interactions drive the dynamics as a response to MGs, and tend to lock the spinor onto a ferromagnetic state, even though depolarization is energetically favoured. Unfortunately this scenario seems to disfavour appearance of beyond mean field effects in our system, which by contrast would reduce the local spin length.  While our configuration is thus probably at the limit of seeing beyond mean-field physics (which could for example be enhanced by working in lower dimensions), we will present soon a new work showing that a very different scenario occurs when the $^{52}$Cr BEC is loaded in a deep optical lattice.

Acknowledgements: We thank O. Gorceix and E. Mar\'echal for important contributions enabling this work, and M. Robert de Saint Vincent for stimulating discussions. The Villetaneuse group acknowledges financial support from Conseil R\'{e}gional d'Ile-de-France under DIM Nano-K / IFRAF, CNRS, Minist\`{e}re de
l'Enseignement Sup\'{e}rieur et de la Recherche within CPER Contract, Universit\'{e} Sorbonne Paris Cit\'{e} (USPC), and the Indo-French Centre for the Promotion
of Advanced Research - CEFIPRA under LORIC5404-1 and PPKC contracts.
AMR acknowledges support by Defense Advanced Research Projects Agency (DARPA, W911NF-16-1-0576 through ARO), NSF grant PHY 1521080, JILA-NSF grant PFC-1125844,  the Air Force Office of Scientific Research and its Multidisciplinary University Research Initiative (AFOSR-MURI).

\newpage
\null
\centerline{\Large \textbf{Supplemental material}}

\vspace{0.5cm}

In this Supplemental Material Section, we describe experimental details concerning the Stern-Gerlach (SG) procedure, measurement of magnetic field gradients, calibration of spin detection measurement. We also discuss why spin exchange interactions provide a feedback mechanism which tends to maintain the ferromagnetic character of the spins during spin dynamics. We finally introduce our hydrodynamic approach for the effect of magnetic gradients. Supplemental experimental data is also provided.

\vspace{0.5cm}

\subsection{Stern Gerlach procedure}

The SG field is in the horizontal plane along a direction x', and has the form: $\vec{B}_{SG}=\left(B_{SG}+bx'\right)\vec{u}_{x'}$, with $b \approx 0.5$ G.cm$^{-1}$. Atoms in different spin states are separated by the associated spin-dependent force. After $t_{TOF}=5$ ms of free fall in presence of the gradient, atoms are imaged through standard absorption imaging, with a 425 nm resonant circularly polarized laser beam. The value of $t_{TOF}$ was chosen in order to have sufficient spatial separation between spin components, while maintaining good signal to noise ratio for imaging (see Fig 1 c)d) of the main article).

\vspace{0.5cm}

\subsection{Measurement of magnetic field gradients}

To measure the magnetic field gradient during spin dynamics, we proceed to a differential measurement, consisting in measuring the ballistic expansion of a BEC prepared for two different $m_s$ states. We let the cloud expand in $\vec{B}_{dyn}$ for 10 ms. We compare the final position of atoms initially in $m_s=3$ with that of atoms initially in $m_s=-3$, by applying or not a frequency sweep across the Larmor frequency $f_L$ before expansion. For $\vec{B}_{dyn}=\left(B_0+\alpha x + \beta y +\gamma z\right)\vec{u}_B+B_1\vec{u}_{\perp}$, with $\left\{\alpha x,\beta y,\gamma z,B_1 \right\}\ll B_0$ in the entire cloud, $\left|\vec{B}\right|\simeq B_0+\alpha x + \beta y +\gamma z$, so that the force experienced by the atom is $gm_S\mu _B\left(\alpha \vec{u}_x + \beta \vec{u}_y +\gamma \vec{u}_z\right)$. To measure $\alpha$, $\beta$ and $\gamma$, we take absorption images along two approximately orthogonal directions.

\vspace{0.5cm}

\subsection{Calibration of imaging system}

The raw values of the populations of the spin components obtained by integration of absorption images do not give absolute values of the populations. This can be for example inferred from the picture shown in Fig 1 c) of the main article. The initial rotation $\theta=\pi /2$ is then expected to lead to a symmetric population distribution ($P_{m_s}=P_{-m_s}$). As evidenced in Fig 1 c) of the paper, there is a growing deficit in population with $m_s$, so that in particular $P_{+3}$ is substantially undervalued.

This systematic effect mostly derives from the fact that the cross section for absorption of resonant light strongly depends on the $m_s$ states, through Clebsch-Gordan coefficients. First, the magnetic field during imaging is not parallel to the imaging beam axis. Second, optical pumping by the imaging beam is not fast enough compared to the 75 $\mu$s duration of the imaging pulse, on the 425 nm $J\rightarrow J+1$ transition, with an intensity of 0.04 the saturation intensity.

Calibration of normalization factors $f_{m_s}$, with $P_{m_s}=f_{m_s}P_{m_s,raw}$, are obtained by comparing measured populations following immediately rotations of $\theta =\pi/4, \pi/2, 3\pi/4,$ and $\pi$, with theoretically expected values. This calibration depends on the magnetic field direction during spin dynamics, as eddy currents do not allow to rapidly set the direction of the magnetic field during imaging. For example for the data in Fig 3 of the main article: $f_{-3}=1 \pm 0.15$, $f_{-2}=1 \pm 0.1$, $f_{-1}=1.4 \pm 0.15 $, $f_{0}=2.25 \pm 0.2 $, $f_{+1}=4.35 \pm 0.45 $, $f_{+2}=5.4 \pm 0.2$, $f_{+3}=6 \pm 2$. Error bars in this calibration are used to compute systematic errors on relative populations.

\vspace{0.5cm}

\subsection{Interaction-driven ferromagnetic  protection in an homogeneous gas}

One of the surprising features revealed by the numerical simulations is that, in the presence of a magnetic field gradient, spin dynamics takes place \textit{while preserving the local ferromagnetic character} of the spinor order parameter. Here we present a short derivation of the conditions for the ferromagnetic protection  for  an homogeneous spinor BEC gas. The results are derived from a linear response analysis of the  Gross-Pitaevskii equation (GPE). For simplicity we set to zero all  subleading spin dependent interactions and only keep $c_1$ (direct exchange ) i.e. $c_2 = c_3 = c_{d d}= 0$. We also set  $V_{\rm trap}=0$ and $B_0=0$. The latter is because the constant part of the magnetic field can be factored out and does not play a role in the dynamics.
The initial condition is a ferromagnetic state along the $x$ direction which we denote as $\Psi_m (x, t = 0) = n \langle m \vert -s_x\rangle$. The exact dynamics generated by the  GPE is rather complex.
However, its short-time limit can be captured quite easily when $b$ is small by postulating a solution of the form :
\begin{displaymath}
 \Psi_m (x, t) = \sqrt{n} \, {\rm e}^{-{\rm i} \, \mu t} [\langle m \vert
 -s_x \rangle + z_m (x, t)],
\end{displaymath}
with $\mu = \bigl( c_0 + s^2 c_1 \bigr) n$ and $\vert z_m \vert \ll 1$
($z_m$ is complex and depends only on $x$), and expand non-linear terms in the
GPE in powers of $z_m$. After some straightforward algebra we arrive at the time-dependent
fractional populations:
\begin{displaymath}
 \frac{p_m (t)}{p_m (0)} = 1 + \frac{\hbar^2 (g \mu_B b)^2}{2 M \nu^3} \zeta
 (\tau) \biggl[ m^2 - \sideset{}{_n} \sum n^2 p_n (0) \biggr]
\end{displaymath}
with $\zeta (\tau) = [\tau^2 - \sin^2 \tau] $, $\tau = \nu t / \hbar$ and $\nu =
c_1 s n$. For times $t< \hbar /c_1 n$, spin dynamics occurs, with a population change $\propto c_1 n$. For large enough spin-dependent interactions, corresponding to $\beta =\frac{\hbar^2 (g \mu_B b)^2}{M \nu^3} \rightarrow 0$, this corresponds to a very small change in populations, which can be in practice neglected. In our experiment,  $\beta < 10^{-6}$. For times $t> \hbar /c_1 n$, while still considering the perturbative regime valid, we only need to keep the quadratic term in $\zeta$ ($\tau^2 \gg \sin^2 \tau$) and the populations remain locked in a ferromagnetic state  when $ (g \mu_B b)^2 t^2 / [ M (c_1 s
n)]<<1$. This condition (also given in the main paper) insures that no population dynamics occurs \textit{for a homogeneous gas} and that the gas remains ferromagnetic.

\subsection{Hydrodynamic approach}

The metastability of the ferromagnetic character of the gas, justified above, provides a way to considerably simplify the theoretical treatment. We phenomenologically \textit{assume} that the BEC always remains locally ferromagnetic, which allows us to derive hydrodynamic equations, which are equivalent to the ferromagnetic Gross-Pitaevskii equation. For a fully polarized BEC, one can find  dynamics equations for the density $\rho$, mass velocity $\textbf{\textit{v}}^{\text{mass}}$, spin velocity $\textbf{\textit{v}}^{\text{spin}}_\mu$ and for the spin density $\textbf{F}$ of component $\mathcal{F}_\mu$ \cite{Kawaguchi2012}. In the following, the dipolar interaction is neglected. The equation for $\mathcal{F}_\mu$ then reads:
\begin{eqnarray}
\frac{\partial \mathcal{F}_\mu}{\partial t} + \nabla.\left[\rho \textbf{\textit{v}}^{\text{spin}}_\mu\right]= \frac{g \mu_B}{\hbar}\left(\textbf{B}_{dyn} \times \textbf{F}  \right)_\mu
\label{eq1}
\end{eqnarray}
where $\mu = (x,y,z)$, $g$ is the Land\'e factor, $\mu_B$ the Bohr magneton, and $\textbf{B}$ the (inhomogeneous) magnetic field. We define the local spin density  $\textbf{\textit{f}} = \textbf{F} /\rho$ and we suppose that $\textbf{\textit{v}}^{\text{mass}}\sim 0$ therefore $\frac{\partial \rho}{\partial t} = - \nabla.\left[\rho \textbf{\textit{v}}^{\text{mass}}\right] \sim 0$ (in accordance with our full numerical simulations of the Gross-Pitaevskii equation). Then Eq.(\ref{eq1}) becomes:
\begin{eqnarray}
\rho \frac{\partial f_\mu}{\partial t} + \nabla.\left[\rho \textbf{\textit{v}}^{\text{spin}}_\mu\right]= \frac{g \mu_B \rho}{\hbar}\left(\textbf{B}_{dyn} \times \textbf{f}  \right)_\mu
\label{hydro_apx_f}
\end{eqnarray}
where
\begin{eqnarray}
\textbf{\textit{v}}^{\text{spin}}_x &= -\frac{\hbar}{2 M s} \left(f_y \nabla f_z - f_z \nabla f_y \right)\\ \nonumber
\textbf{\textit{v}}^{\text{spin}}_y &= -\frac{\hbar}{2 M s} \left(f_z \nabla f_x - f_x \nabla f_z \right)\\ \nonumber
\textbf{\textit{v}}^{\text{spin}}_z &= -\frac{\hbar}{2 M s} \left(f_x \nabla f_y - f_y \nabla f_x \right)
\end{eqnarray}
where $s$ is the spin ($s = 3$ in our case); $M$ is the mass.

Starting from a fully polarized BEC we rotate the spins around the $Oy$ axis with an angle $\theta$. In the presence of an inhomogeneous magnetic field  $\textbf{\textit{B}}_{dyn} =  g \mu_B B_0 \textbf{\textit{e}}_z +  g \mu_B b x \textbf{\textit{e}}_z $, the spin precess around $\textbf{\textit{e}}_z$ (in addition to the homogeneous Larmor precession at frequency  $\omega_L = g \mu_b B_0 x /\hbar$ which factors out of the problem) with a frequency $\omega_b = g \mu_B b x /\hbar$.
We used a Gaussian ansatz for the density  $\rho = \frac{N}{\pi ^{3/2} R^3} \exp \left[- \textbf{\textit{r}}^2/R^2\right]$, with $N$ he number of atoms and $R$ the size of the cloud. We then find:

\begin{eqnarray}
\frac{p_m(t)}{p_m(0)} = \left[ 1 + \left(\frac{g \mu_B b }{2 M R}\right)^2\left( m^2 - \sum_{n} {n}^2  p_{n}(0)  \right)t^4 \right]
\label{simple}
\end{eqnarray}

\begin{figure}
\centering
\includegraphics[width= 8 cm]{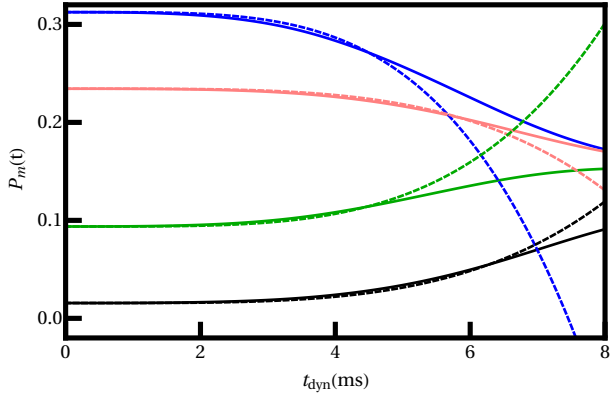}
\caption{Comparison between perturbative model to fourth order, and the full numerical simulation of the Gross-Pitaevskii equation. Populations of the $m_s={-3,-2,-1,0}$ as a function of time (resp. black, green, red, blue). The trap is spherical with a trap frequency of 320 Hz, the atom number is 30 000, and the magnetic gradient is equal to $3.5$ mG.cm$^{-1}$. }
\label{comparison}
\end{figure}

In Fig. \ref{comparison}, we plot the result of Eq.(\ref{simple}) alongside our numerical
simulations of the Gross-Pitaevskii equation without dipole-dipole interactions
for the different fractional populations in states $m_s$. The agreement is
remarkable, confirming the validity of the ferromagnetic approximation and the hydrodynamic approach.

\subsection{Supplemental data}

\begin{figure}
\centering
\includegraphics[width= 8 cm]{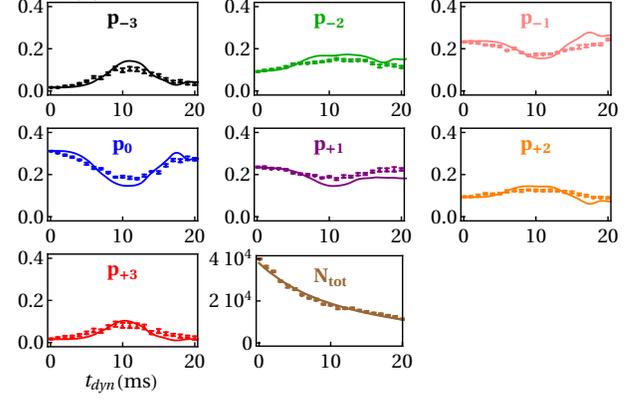}
\caption{Data for smaller gradient than in the main paper. $B = 176$ mG, $\vec{u}_B= \vec{u}_x$, $(b_x,b_y,b_z) = (0 \pm 10, 7 \pm 7, 19.6 \pm 10) $mG.cm$^{-1}$. Full lines
are result of our simulations with no adjustable parameters. }
\label{databis}
\end{figure}

We plot in Fig. \ref{databis} supplemental data, showing the evolution of $p_{m_s}$ for $\theta = \pi/2$ and
a $B_{dyn}$ configuration different than the ones in Fig. 2 a) and Fig. 3 of the main paper. The agreement between experimental data and numerical simulations is as good as the one shown in the paper, confirming the validity of our numerical simulations. We also find that the data shown in the main part of the paper and in this supplemental material all exhibit spin dynamics over the timescale $\left(\frac{2 M R}{g \mu_B b }\right)^{1/2}$ given by Eq.(\ref{simple}).

\vspace{0.5cm}

\end{document}